# Against the Empire

Milan M. Ćirković
*Astronomical Observatory, Volgina 7,
11160 Belgrade, Serbia*
*e-mail:* mcirkovic@aob.bg.ac.yu

**Abstract.** It is argued that the "generic" evolutionary pathway of advanced technological civilizations are more likely to be optimization-driven than expansion-driven, in contrast to the prevailing opinions and attitudes in both future studies on one side and astrobiology/SETI studies on the other. Two toy-models of postbiological evolution of advanced technological civilizations are considered and several arguments supporting the optimization-driven, spatially compact model are briefly discussed. In addition, it is pointed out that there is a subtle contradiction in most of the tech-optimist and transhumanist accounts of future human/alien civilizations' motivations in its postbiological stages. This may have important ramifications for both practical SETI projects and the future (of humanity) studies.

**Keywords:** extraterrestrial intelligence – SETI – Fermi's paradox – future studies – transhumanism – imperialism – interstellar colonization

> *Turning and turning in a widening gyre*
> *The falcon cannot hear the falconer...*
>
> W. B. Yeats (1920)

**1. Evolutionary pathways of intelligent communities and ATCs**

In this paper, I shall present some arguments to the effect that there is a subtle contradiction in most accounts of motivations and future evolution of advanced intelligent communities of both (post)human and extraterrestrial origin. It is clear that the same mode of reasoning could be applied to future human civilization and to hypothetical extraterrestrial civilizations, thus the future human or posthuman civilization serves as the prototype in this section (and the relevant applications to extraterrestrial civilizations will be clear from subsequent sections). The term "posthuman" denotes possible descendants of the present-day humans whose basic capacities so radically exceed those of present humans as to be no longer unambiguously human by our current standards; I use it here without any value judgement, since the term is used in both desirable (e.g. [1]) and undesirable (e.g. [2]) contexts. Research program dealing with ramifications, promises, and potential dangers of technologies that will enable us to overcome fundamental human limitations and possibly achieve transition to posthuman stage of evolution is often called transhumanism; although the term can have other meanings [1-5]. While most transhumanists seem to agree that the posthuman civilization will be largely or even exclusively postbiological – consisting, say, of uploads or Moravec-like robots or a single hive-mind – they are far less ready to commit themselves to abandoning of motivations characteristic for biological actors (and, even more, products of neodarwinian evolution and adaptationist paradigm). This is even stranger when one takes into account the fact that most, if not all, technological optimists are enthusiastic supporters of not only any naturalistic, but specifically neodarwinian, evolutionary account of mankind's origin and even the sociobiological origin of human culture and technological civilization. The situation seems to be as follows: if we agree that specific biological motivations have been a determining factor in the biological (human) phase of the history of our species, it would be only reasonable to argue that, with the transition to the postbiological (posthuman) phase, the old biological impetuses and motivations will become largely irrelevant. Paradoxically, it is rare to encounter such attitude in tech-optimists/transhumanist circles; in general, the predominant view is that the posthumanity will enable faster, better, larger, etc. steps toward achieving the same **old**, biological, Darwinian aims and goals. In other words, just new means toward old ends. I hereby argue that such view is old-fashioned, illogical and ultimately untenable. Rejecting it could throw some new light on issues in both future studies, as well as the discussions of advanced extraterrestrial civilizations and ongoing SETI projects.

One reason why it is so difficult to dissociate oneself from the biological imperatives is the too-strong impression made by successes of the neodarwinian theory of evolution in eyes of most educated people, including all transhumanists. However, although the very word "transhumanism" was coined by Sir Julian Huxley, one of the founding fathers of the Modern Synthesis in life sciences [3]. Huxley himself was completely clear that the natural selection has nothing to do in the future transhumanist stage of human evolution. Other mechanisms, characterizing cultural, technological and other modes of evolution will become dominant and bring with them the whole new array of issues and concerns. The same sentiment is occasionally found at the other side of the divide between natural sciences and humanities; for instance, Toynbee [6] in an essay appropriately entitled "The Acceleration of Human History" speaks about "metabiological" phase of human evolution we are entering, which could be construed as an intermediate stage between biological and postbiological epochs, where some biological imperatives remain, but others are muted. Here, I try to investigate the logical consequences of following the same pathway. Some of the relevant issues have been prefigured in a poetic and philosophical form by Teilhard de Chardin [7], most notably his insistence on accessibility of knowledge and rapidity of communication as preconditions for a new phase of evolution of humanity. Notable recent work of John Smart reaches very similar conclusions, while starting from somewhat different premises [8].

In this paper the concept of an advanced technological civilization (henceforth ATC) from the study of Ćirković & Bradbury [9] is retained. ATCs are advanced outcomes of cultural evolution which are immune to most existential risks, barring possible universe-destroying ones (e.g., vacuum phase transition) and which have reached sufficient capacities for manipulating surrounding physical universe on large scale and with almost arbitrary precision. Thus, an ATC would reach the Type II of Kardashev's classification [10], based on the energy utilization; that is, an ATC would use all energy resources of its domicile planetary system. However, it is one of the purposes of the present paper to criticize the applicability of Kardashev's classification, which I believe is of very limited value in the real SETI effort and is partially misleading. ATCs as discussed here have some of the general trademarks of the posthuman civilization envisaged by diverse authors such as Stapledon [11], Huxley, Bostrom [12,13] or Kurzweil [14,15]. In other word, posthuman civilization would be a realization of ATC in the specific environment of the Solar System. This does not automatically mean that all characteristics often cited in relation to the concept of posthumanity need to apply (or even are reasonable to expect).

For instance, one of the characteristics of posthuman society according to Bostrom is a large population of the order of $10^{12}$ persons [13]; for mammals evolved in the Darwinian manner in the specific terrestrial condition this may indeed be a useful benchmark. It would be too narrow, however, to speculate that extraterrestrial intelligent species cannot achieve the same or greater degree of control over their environment with either much smaller (in the "hive mind" scenarios even equal to unity!) or much larger (if individually less capable) population. While some discussion of these and related issues has been present in the form of science-fictional discourse since Stapledon [11], it is my contention that it is high time for the debate to enter the new level of discussion, from both astrobiological and future-studies points of view.

The present discussion has dual aspects of pertaining to both future (post)human civilization and those alien civilizations which have already achieved ATC status elsewhere in the Galaxy. In this sense the present paper extends the preliminary discussion of the relevance of SETI research for transhumanism [16]. Some of the auxiliary arguments are clearly more relevant to the future of humanity than to extraterrestrial ATCs, and are discussed here as such, but the bulk of the discussion can be considered equivocal in this respect. As far as transcending our anthropomorphic biases is concerned, we may speak of these as particular future historical trajectories in the conceivable parameter space without specifying the parameters or even claim that we are aware of all of them. Of course, since there is no clear standard for anthropocentrism, we cannot ever claim that our concepts are free of that bias. The best we can strive for is to acknowledge the presence of a bias and try to isolate points in which the discussion needs to be generalized further.

This is related to another strong misconception: the idea that there is a "see-saw" tension between the optimism for future of human life and the optimism for life in general cosmic context and that we should feel "depressed" at prospects of finding extraterrestrial life and intelligence [17,18]. This misconception depends on (A) an incomplete and superficial reading of Fermi's paradox (more on that in §8 below), and (B) the highly questionable understanding of evolutionary "success", the one which cannot logically be linked to postbiological evolution; e.g., Hanson writes [17]:

> Similarly, technological "optimists" have taken standard economic trends and our standard understanding of evolutionary processes to argue the plausibility... that our descendants have a decent chance of colonizing our solar system and then, with increasingly fast and reliable technologies of space travel, colonizing other stars and galaxies. If so, our descendants have a foreseeable chance of reaching such an explosive point within a cosmologically short time (say a million years)...

But why should they, especially when one considers many additional dimensions the relevant evolutionary parameter space will have at that point? In some other context, with possible evolutionary subspace more tightly constrained, Hanson's reasoning would have been valid; in a larger space of possibilities, it is only a special case, very far from the "manifest destiny" this way of reasoning implies. There is no proof that "colonizing other stars and galaxies" constitutes anything more than a subset of zero-measure trajectories in the evolutionary space – and there are indications to the contrary which will be considered in the rest of this paper.

## 2. Two basic models

Two basic models listed below are undoubtedly oversimplified and extreme, but their consideration will enable easier discussion of more complex and more realistic models which will contain a mixture of these two prototypes.

### 2.1. "The Empire-State"

This is the classical "expand-and-colonize" model. Limits to growth are soft and to be easily overcome. Expansion is virtually unlimited, even when faced with the limits of physical eschatology [19-21]. Typical ATC spreads out among the stars, utilizing resources in a large spatial volume, and increasing the number of observers (or observer-moments, see Ref. 22) indefinitely or at least for astrophysically relevant duration. This model essentially corresponds to Kardashev's Type III civilizations or the ascent towards Type III analogs.

### 2.2. "The City-State"

This is the "Olympian perfection" model. Limits to expansion and growth are hard and the optimization of all activities, most notably computation is the existential imperative. Moreover, some of the limits are internalized, i.e. an advanced civilization willingly imposes some of the limits on the expansion. Expansion beyond some critical value will tend to undermine efficiency, due both to latency, bandwidth and noise problems (cf. [23]). A typical ATC utilizes resources of its domicile planetary system and – possibly, but not really necessarily – other nearby planetary systems, conceivably by bringing resources back home instead of truly colonizing them. Instead, the future evolution of an ATC will be more and more optimization-driven. In the limit of very long timescales characterizing ATC planning and strategies, it may lead to relocation in physical space (for a plausible reason for such migrations, see Ref. 9). However, this need not mean relinquishing of the basic optimization-driven model, just reinstatiating it.

## 3. Postbiological evolution

Clearly, we know very little at present about the modes of postbiological evolution. However, even a minimal framework derived from the very meaning of "postbiological" can still be very useful. Notably, the transition to postbiological phase obviates most, if not all, biological motivations. The very definition of ecology and the relevant ecological needs and imperatives changes, leading to significant changes in other fields which have been traditionally linked to the evolutionary processes.

As an example, the imperative for filling the complete ecological niche in order to maximize one's survival chances and decrease the amount of biotic competition is an essentially biological part of motivation for any species, including present-day humans. (Here I do not presuppose that motivation is a product of consciousness, rather than, say, adaptive strategy for fitness optimization.) It would be hard to deny that this circumstance has played a significant role in colonization of the surface of the Earth. But expanding and filling the ecological niches are not the intrinsic property of life or intelligence – they are just consequences of the predominant evolutionary mechanism, i.e. natural selection [24-27]. It seems logically possible to imagine a situation in which some other mechanism of evolutionary change, like the Lamarckian inheritance or genetic drift, could dominate and prompt different types of behaviour. The same applies for the desire to procreate, leave many children and enable more competitive transmission of one's genes to future generation is linked with the very basics of the Darwinian evolution. Postbiological civilization is quite unlikely to retain anything like the genetic lottery when the creation of new generations is concerned. In addition, the easiness of producing and retaining copies of postbiological organisms in the digital substrate are likely to dramatically change the meaning of terms such as "maturation", "adulthood", "parenthood", "kin", etc. Thus, we need to make an additional step symbolically represented as the analogy

> biological evolution $\rightarrow$ postbiological evolution
> sociobiology $\rightarrow$ "post-sociobiology"

Clearly, we need much more research and thinking in order to establish what exactly could "post-sociobiology" be, but as a provocation we may suppose that it will deal with "stable ingredients" (to use the expression of Arnold Toynbee; see Ref. 6) of postbiological development. In the case of (post)human evolution, one may argue that this will encompass "posthuman nature" in the same manner as authors like Fukuyama [2] invoke "human nature" as an explanatory device. It is very hard to imagine such a dramatic change – but we still ought to think as hard as possible about its outcomes since, among other things, some very early decisions can have long-reaching consequences [28].

In the postbiological context, some of the common criticisms of the city-state model are obviously obviated (e.g., "life tends to spread"), while the others are undermined in a more subtle way (e.g., "universe may still contain existential dangers"). While limitations imposed on this model are serious, its acceptance does not mean, as some authors have misconstrued it, the loss of interest for the external physical world ("playing computer games" or "sitting home and surfing the net" as has been derisively commented upon). On the contrary, information-centric view which emphasizes the need for optimized information processing will lead to very careful and detailed monitoring of the Galaxy by city-state ATCs, for both research and practical reasons [29,30]. It is as appropriate to expect such an ATC to develop "big eyes" and "big ears" (and an array of nanotechnology-based interstellar probes and monitors), as it was common for the city-states of ancient Greece to maintain complex networks of agents and scouts outside of their territories (cf. Ref. 31). Fictional posthuman polises in Egan's *Diaspora* pursue different interests and agendas, some which are undoubtedly "introvert", but the others showing clear interest in the external world (even more, the views of the latter actually are shown – at least for the sake of the drama – to be vital for the survival of humanity). For instance, a key piece of "post-sociobiology" substituting for the Wilsonian epigenetic rules could be what the great historian of science Steven J. Dick has dubbed the *Intelligence Principle* [32]:

> In sorting priorities, I adopt… the central principle of cultural evolution, which I refer to as the Intelligence Principle: the maintenance, improvement and perpetuation of knowledge and intelligence is the central driving force of cultural evolution, and that to the extent intelligence can be improved, it will be improved.

In the rest of this paper, I present some of the arguments against the expansionist model upon the assumed postbiological background.

## 4. Against the Empire: lack of motivation

Let us suppose that the very definition of ATC (subject on which vagueness is mandatory, lacking any form of certain knowledge) includes transition to postbiological stage, as many serious future-oriented thinkers have suggested [12,13,15,32-34]. It is not important for the present purposes which of the proposed scenarios ("uploads", "robots", etc.) of the transition and its next stage are correct, since they are all just simplified scenarios of an inherently complex and heterogeneous evolutionary process. One thing follows logically: within the ATC context, traditional biological imperatives, like the survival until the reproduction age, leaving as numerous and as healthy progeny as possible, protection of infants, various forms of biologically determined social structures will become marginal, if not entirely extinct as valid motivations for individual and group actions. Let us, for the sake of elaborated example, consider the society of uploaded minds living in virtual cities of Greg Egan's *Diaspora* [35] – apart from some very general energy requirements, making copies of one's mind and even sending some or all of them to intergalactic trips (with subsequent merging of willing copies) is cheap and uninfluenced by any biological imperative whatsoever; the galaxy is simply large and they are expanding freely, in many different ways with no clear hierarchy of approaches. There is no genetic heritage to be passed on, no truly threatening environment to exert selection pressure, no necessity to retain biologically determined sexual characters, no biotic competition, no kin selection, no pressure on (digital) ecological boundaries, no minimal viable populations. (The global Galactic catastrophe revealed in the course of Egan's novel clearly is such an external threat, but it serves predominantly as a dramatic device, being astrophysically untenable, and we can neglect such extremes in the present context.) Since all these and many others biological phenomena have been quite certainly underlying the human (and presumably other alien) life and culture until some particular point in its evolution, it is clear that in the ATC context there is no biological explanation/justification for historical phenomena usually justified in this manner: expansion, colonization, pushing the boundaries outward. Ironically, the very acceptance of sociobiological and evolutionary psychological explanations of human cultural and historical behaviour (common in today's transhumanist/future studies circles) forces us to accept "the other side of the coin": that with the postbiological transition, most if not all motivation for the "empire state" model evaporates. In particular,

1. Molecular nanotechnology and the related developments will obviate the economic need for imperial-style expansion, since the efficiency of utilization of resources will dramatically increase. This would prompt a sort of vicarious "silicon colonization" of economically significant cosmic resources [28], rather than any direct human presence.
2. Occasional other reasons have been considered as explanations for historical expansionism, notably religious fervour and the feeling of moral superiority (e.g. Ref. 36). Both are unlikely to play a significant role either in future of humanity or in functioning of extraterrestrial ATCs.

On the other hand, what is the real criterion of a civilization's success or failure? In the limit of long future timescales, we seek those features which are truly fundamental and important. Obviously, there can be much reasonable disagreement on this issue, which intrigued some of the greatest thinkers of mankind, from Plato to Jared Diamond.

Clearly, the question has a tautological aspect, since some of the terms involved in phrasing the question itself – like the term "success" – have meaning which are at least in part civilization-dependent. Our inquiry is doomed if the entire meaning of these are civilization-dependent; the history of the problem itself testifies that we can hope to find some more-or-less universal construal. One of the plausible criteria for success would be capacity of an agent for making the substrate of our environment more and more malleable to agent's wishes and desires. Since no amount of technological and social progress will make physical substrate infinitely malleable, shaping digital substrate which obviously can be infinitely malleable will certainly be more appealing to both communities and individuals tending to perfectionism. On the other hand, global accessibility of information is easy to achieve without widespread physical presence (as we have discussed in relation to miniaturization and "big eyes" above). All this clearly favours the "city state" model, as far as the motivation is concerned.

It has been claimed in the classical SETI literature that the interstellar migrations will be forced by the natural course of stellar evolution [37]. However, even this "attenuated" expansionism – delayed by on the order of $10^9$ years – is actually unnecessary, since naturally occurring thermonuclear fusion in stars is extremely inefficient energy source, converting less than 1% of the total stellar mass into potentially useable energy. Much deeper (by at least an order of magnitude) reservoir of useful energy is contained in the gravitational field of a stellar remnant (white dwarf, neutron star or black hole), even without already envisaged stellar engineering [38,39]. Highly optimized civilization will be able to prolong utilization of its astrophysically local resources to truly cosmological timescales. The consequences for our conventional (that is, predominantly empire-state) view of ATCs have been encapsulated in an interesting paper by Beech [40]:

> [A] star can only burn hydrogen for a finite time, and it is probably safe to suppose that a civilisation capable of engineering the condition of their parent star is also capable of initiating a programme of interstellar exploration. Should they embark on such a programme of exploration it is suggested that they will do so, however, *by choice rather than by necessitated practicality*. [emphasis M. M. Ć.]

In brief, the often-quoted cliché that life fills all available niches is clearly *non sequitur* in the relevant context; thus, interstellar colonial expansion should not be a default hypothesis, which it sadly is in most SETI-related and far-future-related discourses thus far.

## 5. Against the Empire: feasibility costs

It is not clear to which extent space colonization efforts are both feasible and profitable. Obviously, this issue depends on the technological and societal details of future organization of economy, technology and research. However, it is clearly not likely that the cost of interstellar expansion will ever be low – at least as long as our current understanding of physical laws is basically correct and no "shortcuts" (provided, for instance, by traversable wormholes) exist.

Historical experience of human colonization of Earth by (mainly European) military powers offers an ambiguous record, but on the balance it seems that costs outweigh the benefits. In the case of first colonial powers, Spain and Portugal, this is abundantly clear; some of the historians of imperialism claim that the same holds for all other imperial powers (e.g., Refs. 41-43).

Of course, costs rise astronomically if – as one may expect in the realistic case – some colonization sites are already occupied and the prospective colonizer must expand through war and conquest instead through unopposed colonization. Costs increase exponentially, if active opposition to the expansion is to be expected. In an "arms race" situation, real efficiency of utilization of resources dramatically decreases, since war-related expenses (even if we disregard other disastrous consequences of militarism) will certainly decrease the amount of resources allocated to achieving real goals of a civilization. Expansionist civilizations will create an expanding front of industrial activity surrounding largely exhausted volume of space. Some agents will probably be left behind to oversee the usage of long-term resources and macro-engineering projects on the longer timescales. But, in general, the empire-state model leads to dramatic "burning of cosmic commons" scenario of Hanson [44]. Since we are clearly aware of such a possibility so early in our own development, Hanson's scenario is almost a *reductio ad absurdum* for expansionism and imperative to avoid such a situation. Since opportunity costs increase with the delay of strategic decisions in the long-term perspective [28], it seems clear that ATCs are likely to commit to a different model of evolutionary trajectory early in their histories. Again, it does not need to be our – excessively simplified, it is worth repeating – city-state model, but is likely to contain some of its elements. As pointed out by Parkinson [77], one may even envisage a sort of "containment" of would-be empire-builders by their more efficient city-state Galactic neighbours.

Another relevant issue borders on the obvious: truly efficient system is rather hard to observe. Consider, for instance, light pollution in Earth's urban areas: most proposals for light pollution reduction are motivated not only by the lack of a romantic view of stars and other celestial bodies or by selfish concerns of astronomers, but by a reasonable conclusion that strong light pollution must mean low efficiency of street lamps and other photon sources. Light pollution is necessarily a waste of resources. On the other hand, it is exactly light pollution which makes Earth *visibly* inhabited by a technological species (when the night side is observed from afar). The same applies to other forms of energy: radiowaves, microwaves, etc. Thus – and especially in the context of the conventionally understood SETI "listening" – one is *likelier to detect a wasteful than a truly efficient civilization from afar*. If we postulate that ATCs present the summit of efficiency, where every single photon counts, than something close to invisibility – irrespective of arbitrary things such as cultural wishes or phobias – is a logical consequence of the optimization drive.

**6. Against the Empire: interstellar ethics**

If many locales in the Galaxy are inhabited, even by low-level lifeforms (in accordance, for instance, with controversial, but popular "rare Earth" hypothesis; see [45]), the problem of planetary contamination gets much wider and serious aspects [46,47]. NASA recently acknowledged the validity of these questions by changing the terminal trajectory of its "Galileo" spacecraft to avoid even a remote possibility of contamination of a hypothetical ecosystem on Jupiter's moon Europa by organisms of terrestrial origin. It is clear that there is no possibility to retain pristine planetary biospheres in face of widespread colonization and economic exploitation. On the contrary, supplanting local ecosystems with the imported ones is quite plausibly a *sine qua non* of any colonizing endeavour. Even if local species are preserved in isolated locales, the wholeness of their habitat will be irrevocably and irreparably destroyed.

Even our extremely limited terrestrial experience indicates serious ethical concerns about this. Do our human or posthuman descendants possess any moral right to influence (not to mention supplant or destroy) alien biospheres on other worlds? One could argue that in the case of utter necessity of survival this can be adequately justified, but, as I try to argue here, the expansionist model cannot become utter necessity as long as any other model is viable. In the city-state model, this ethical dilemma simply does not exist (under the very weak assumption that presumably extremely miniaturized research probes – "eyes and ears" – can be contamination-proof).

Of course, if it turns out that distant resources are controlled by other intelligent species (which will certainly be possible to check from afar or with miniaturized probes), expanding to take control of them will constitute an act of aggression. Such acts will be even more difficult to ethically justify in the cosmic context of far future than analogous acts of resource-grabbing are in the human world of today.

**7. Against the Empire: interstellar politics**

In classical Greece, the instances of imperialism – notably the Athenian Empire of 5th century BC and its subsequent Spartan, Theban and Macedonian emulations – were traditionally and constantly compared to a tyranny, a word hateful to the Greeks [36,48,49]. It was clear to the enlightened Hellenes that the desire to dominate other lands and people is the manifestation of the same underlying causes as the malicious desire to dominate one's own compatriots. Hopefully, the aversion is shared by us and will be shared by future generations – and for a good practical reason.

If we follow the taxonomy of existential risks by Bostrom [18], only relevant risks for an ATC are those following from long-term processes such as dysgenic evolution or consequences of internal social disaster, such as destructive internal strife or totalitarianism (similar conclusion has been reached by other recent authors interested in global risks; e.g. Ref. 50). Out of these, the one which seems to have most staying power is the possibility of a global totalitarianism, which may actually increase as we approach posthuman stage of our development [51]. It is reasonable to assume that the threat of totalitarianism of some kind is present in any form of community of intelligent beings. As far as extraterrestrial civilizations are concerned, totalitarianism is likely to drastically decrease the contact cross-section and if it is generically likelier than we usually assume, it may explain part of the "Great Silence" problem (see §8 below).

If global totalitarianism remains viable even for ATCs, it presents a problem for both our developmental models. However, it seems that there is a weak imbalance here and that the empire-state model is somewhat easier to subvert – the totalitarian temptation is much harder to resist in conditions where massive military/colonization forces are in existence and thus prone to be misused against state's own citizens. As the classical critic of Victorian imperialism melodramatically asked [52]:

> Is it possible for a federation of civilised States to maintain the force required to keep order in the world without abusing her power by political and economic parasitism?

An extremely interesting SF-rendition of this dichotomy is given in *The Golden Age* trilogy by John C. Wright (*The Golden Age*, *The Phoenix Exultant* and *The Golden Transcendence*; Refs. 53-55) where the two future human civilizations we encounter (the "Golden Oekumene" and the "Dark Oekumene") roughly correspond to our two models above. The Golden Oekumene is, by the time of story's opening, almost perfect realization of the city-state model – comprising most of the Solar System, utilizing stellar uplifting on the Sun and other advanced technology to create a peaceful paradise of unprecented freedom and cultural diversity.

In an ironic overstretch, the military forces of the Golden Oekumene are reduced to a single person – and with perfectly good and rational reasons. Although the resolution of the trilogy transforms the city-state-like Golden Oekumene into an expansionist and militaristic force, it is rather doubtful whether such a resolution is really necessary, much less whether it is desirable in the general context. Important particulars here include obviously totalitarian character of the primary expansionist power (the Dark Oekumene), as well as *de facto* subversion of the decentralized, democratic and peaceful government of the Golden Oekumene, as well as the emergence of the personality-cult mentality.

**8. Against the Empire: astronomical observations and Fermi's paradox**

Quite simply, the astronomical case against the empire-state model is that we haven't noticed any interstellar empire thus far, although with rather weak additional assumptions they should have already be there, encompassing, perhaps, Earth and the Solar System as well (traditional Fermi's paradox or the "Great Silence" problem; [56,57]). Recent astrobiological [9,58], as well as cosmological [59] research makes the problem – or, rather, tightly interconnected set of problems – significantly more serious than hitherto assumed. One important point to keep in mind is the result of Lineweaver (Ref. 60; expanded and reinforced in Ref. 58), indicating that the median age of terrestrial planets in the Milky Way is about 1.8 Gyr greater than the age of Earth and the Solar System. By Copernican assumption, the median age of the technological civilizations should be greater than the age of human civilization by the same amount. The vastness of this interval – and, moreover, we are interested in those habitable planets older than this median value! – indicates that one or more processes must suppress observability of extraterrestrial intelligent communities.

In principle, three broad classes of answers have been advanced which could explain the "Great Silence" in a naturalistic manner. Either (1) we are the only intelligent species in the Galaxy, or (2) it is impossible/unfeasible/too early to create interstellar empires, or (3) they are in fact here, either hiding or conveniently manipulating our observations. Let us immediately discard the third class as too speculative (it includes well-known "Zoo," "Interdict," and "Planetarium" scenarios; Refs. 61-63; the "simulation argument" of Bostrom [64] could also be related to this kind of answer). The first option is fashionably construed as the "rare Earth" hypothesis [45] and has become quite widespread in astrobiology, in spite of occasional strong criticisms (e.g. Ref. 65). In fact, it goes back to the (in)famous argument of Carter [66] which attempts to use the anthropic reasoning to argue for our uniqueness. Note, however, that even the most extreme "rare Earthers" readily admit that simple lifeforms are likely to be ubiquitous throughout the Milky Way; they only take the probability of transition to complex life and tool-making intelligence to be negligible.

The alternative is that there is either physical (supervolcanism, gamma-ray bursts, etc.) or social (exterminating nuclear warfare, global totalitarianism) reason for attenuation of the formation of large and by definition observable extraterrestrial civilizations. For many reasons, the physical subset of these neocatastrophic hypotheses is preferable, especially if the catastrophic events are temporally correlated over the entire Galaxy or its large parts [67]. Similar to this is the idea that there are large phases of stagnation alternating with the expansion of ATCs through the Galaxy; a numerical model for expansion/stagnation equilibrium compatible with Fermi's paradox has been presented by Gros [68]. Still, all such solutions are suspicious since it is unclear how non-exclusive they really are in view of the probably large number of habitable sites in the disk of our Galaxy.

However, both these viable approaches resolve the "Great Silence" problem in the same manner: they assume that ATCs would be easily detectable if they are out there (at least in the Milky Way). In other words, they uncritically apply the usually assumed model of expanding "colonial empire" from human history. In the words of Stanislaw Lem in his famous and in many ways prescient futurological treatise *Summa Technologiae*, we uncritically assume that the generic history of a technological civilization is "orthoevolutionary" [69]: that its contact cross-section monotonically increases. It is not clear whether we can reasonably talk about observability of extraterrestrial ATCs with our current technology if their evolutionary pathway is at all similar to the city-state model outlined above. Ćirković & Bradbury [9] discuss this issue at some length and offer tentative proposals. For the present purposes, it is enough to mention that it is not the existence of extraterrestrial ATCs *per se*, but the assumption of the empire-state model which confronts us with the gravest forms of Fermi's paradox. On the other hand, compact, highly efficient city-state ATCs will easily pass unnoticed even by much more advanced SETI equipment, especially if located near the Galactic rim or at other remote locations. Parkinson's "containment" scenario [77] offers a different rationale for predominance of the city-state over the empire-state model, resulting in the same observed dearth of interstellar empires.

The circumstantial evidence for this view comes from extragalactic SETI. It has been correctly argued that large, galaxy-spanning civilizations would be visible even over intergalactic distances, and that even if we are the only technological species in the Milky Way, we should have observed large parts of nearby galaxies (like M31 or the Large Magellanic Cloud) transformed into Dyson shells, Matrioshka brains, computronium, luxurious mansions or whatever artificial form ATCs prefer. Now, "rare Earthers" could take their version to the extreme and to argue that we are unique not only in our Galaxy, but on a wider stage (for an early view of this kind, see Ref. 70). However, taking into account the architecture of the universe, it takes an additional great leap of scepticism to argue that we are not just alone in the Milky Way but alone in the Local Supercluster – or even in the visible universe comprising $\sim 10^{11}$ galaxies. And yet, Type III civilizations should have been observed in other galaxies and preliminary observational research shows that they are not [71].

Of course, it is possible that there are physical reasons preventing formation of Type III civilizations *by this date* in cosmic history. I have proposed such an alternative scenario [72] in reply to an interesting philosophical puzzle of Ken Olum [59]. In such approach, although Type III civilizations are in principle possible, there has been only insufficient time since the galaxy formation epoch for one to actually emerge. Both physical laws and our anthropic reasoning remain unscathed in such a scenario. However, this temporal delay does not exclude our present working hypothesis that even at any future time Type III civilization will be less feasible and therefore less probable than other pathways of cultural evolution.

Thus, the astronomical argument is not conclusive in itself. And yet, in conjunction with the other lines of thought, and considering the direction of advances in astrobiology in recent years, it offers further corroboration to the assertion that building of the interstellar empires (Kardashev's Type III civilizations) is at least extremely difficult and unlikely. Thus, most of the talk about "Great Filter" (e.g. in Refs. 17,18) is, in fact, misplaced, since it is filtering out only a subsection – and, I wish to argue here, rather small subsection at that – of possible evolutionary pathways: those leading to the empire-state civilizations. The absence of the latter from our cosmological neighbourhood gives a weak probabilistic support for the prevalence of the city-state model, *ceteris paribus*.

**9. Against the Empire: history of humanity**

Admittedly weakest arguments against the "empire-state" model come from considerations of the human history on Earth thus far. (Clearly, though, the same weakness applies to opposite conclusions often drawn from the historical experience.) It is too often forgotten – both among SETI proponents, as well as the contact pessimists – that the colonial expansion has been an exception, rather than the rule in human history so far; our Western-centric attitude should not blind us into accepting a wrong model for civilizational behavior. Countless city-states, be they in ancient Greece, pre-Aryan India, Babylonia, Arabia of Prophet's time, medieval Italy, Germany or Russia, pre-Incan Andes or Mayan Mexico, have all together much longer and stronger traditions than imperial powers, of which there are no more than two dozen examples altogether, from Assyria to the contemporary USA. Even in the cases where cities and other smaller organizational units have been peacefully or otherwise incorporated into a larger whole, this was often regarded as optimization of resources and management, and clear limits to growth have been set in advance; examples in this respect range from the Achaean or Aetolian Leagues, to Hansa, to Swiss Confederation, to China after Ch'in unification. It is exactly this understanding of limits (or resources and communication) which made robust longevity of civilizations like the Chinese, or organizations like the Roman Catholic Church so prominent in the human history so far. *Vice versa*, it was disregard for these limits which at least contributed to downfalls of all historical empires. Current ascendancy of large states and quasi-imperial nations should not blind us through insidious observation-selection effects and biases to the fact that such state of affairs is largely atypical (e.g., Ref. 73).

David Hume encapsulated the Enlightenment repulsion toward grand imperialist unification projects in the famous passage from *Of the Rise and Progress of the Arts and Sciences* [74]:

> Nothing is more favourable to the rise of politeness and learning than a number of neighbouring and independent states connected together by commerce and policy. The emulation which naturally arises among those neighbouring state is an obvious source of improvement; but what would I chiefly insist on is the stop which such limited territories give both to power and to authority... Where a number of neighbouring states have a great intercourse of arts and commerce, their mutual jealousy keeps them from receiving too lightly the law from each other in matters of taste and reasoning, and makes them examine every work of art with the greatest care and accuracy. The contagion of popular opinion spreads not so easily from one place to another. It readily receives a check in some state or other, where it concurs not with the prevailing prejudices.

Communication theorist and fiction author Paul Levinson shows, in his recent book *RealSpace*, that there is already a strong and dangerous misbalance between the human capacities in different technological fields [75]. The book surveys the history of two major human activities, communication and transportation ("talking and walking"), and points out the possibly fatal imbalance brought about by explosive development of digital technology on one side, and gradual marginalization of space travel amidst decreased public and leading intellectual interest on the other. Technologies of communication and transportation, which have been developed almost in parallel (printing and sailing, radio and automobile, television and passenger airplane), are now seriously imbalanced, since there is no new frontier in the realm of travel and transportation which matches the digital revolution in the age of the Web.

It is somewhat ironic that Levinson actually makes good case for increasing the investments in space travel and its general visibility and importance in human culture. However, this still does not make him a proponent of the empire state model – it is reasonable to assume that, for instance, after the Solar System is effectively technologized, most of the rationale for the long-range space travel will be dissolved (for instance, the fear of the planet-wide ecological cataclysm).

## 10. Conclusions

As much as our understanding of the conditions and social dynamics of ATCs is negligible, some of the general issues may and should be speculated upon even at the present-day stage. This is relevant for both the future of humanity and for assessing our own SETI-projects thus far. In brief, the discussion in this paper can be summarized as follows:

- The belief that an intelligent community which survives all catastrophic risks and develops advanced technology will inexorably or even likely colonize the Galaxy is an unsupported dogma essentially equivalent to the belief in Fukuyama's mystical "Factor X" [2] and stemming from the same naive organicism.
- Although the real set of postbiological evolutionary pathways is likely to be immensely more complex, it still makes more sense to discuss it in the framework of the compact city-state model rather than conventionally assumed empire-state model.
- Astronomical observations confirm that there are no star-powered Kardashev's Type III civilizations in our cosmological neighbourhood, which is most plausibly explained by assuming that the measure of postbiological evolutionary pathways leading to such galactic empires is very small or vanishing.
- Transhumanist and future studies should devote more attention to the relationship between efficiency of resource utilization and the character of cultural evolution (including the observability of a particularly evolving model civilization from afar).

Since our astrophysical knowledge clearly precludes infinite expansion [76], it is certainly worthwhile to investigate, at least in the most general terms, logical alternatives to it. I argue that even finite expansion makes sense only within clear limits, delineated by astrophysics, postbiological evolution and even political and moral considerations. These limits do not include civilizations of Kardashev's Type III. Thus, their absence from our astronomical observations is neither good nor bad sign as far as the future of humanity is concerned – the very concept of Type III civilization is irrelevant concept in the first place. There is no need for a frantic search for the "Great Filter", much less for expressing pessimism vis-à-vis astrobiological mission of search for life and intelligence in the universe.

**Acknowledgements**. Insightful comments of two anonymous referees enormously helped in improving a previous version of this manuscript. This work has been partially supported by the Ministry of Science of the Republic of Serbia through the project ON146012. Useful discussions with Anders Sandberg, Nick Bostrom, Irena Diklić, Robert J. Bradbury, George Dvorsky, Karl Schroeder, Cosma R. Shalizi, Slobodan Popović, and Robin Hanson are also hereby acknowledged. This is an opportunity to thank *KoBSON* Consortium of Serbian libraries, which enabled at least partial overcoming of the gap in obtaining the scientific literature during the tragic 1990s.

**References**

[1.] Bostrom, N. 2005, "In Defence of Posthuman Dignity," *Bioethics* **19**, 202-214.
[2.] Fukuyama, F. 2002, *Our Posthuman Future: Consequences of the Biotechnology Revolution* (Farrar, Straus & Giroux, New York).
[3.] Huxley, J. 1957, "Transhumanism," in *New Bottles for New Vine* (Chatto & Windus, London), pp. 13–17.
[4.] Bostrom, N. 2003, "The Transhumanist FAQ" (available at http://www.transhumanism.org/resources/faq.html).
[5.] Hughes, J. 2004, *Citizen Cyborg* (Westview Press, Cambridge).
[6.] Toynbee, A. 1966, *Change and Habit* (Oxford University Press, Oxford).
[7.] Teilhard de Chardin, P. 1964, *The future of man* (Harper & Row, New York).
[8.] Smart, J. 2007, "Answering the Fermi Paradox: Exploring the Mechanisms of Universal Transcension," (preprint at http://accelerating.org/articles/answeringfermiparadox.html.)
[9.] Ćirković, M. M. and Bradbury, R. J. 2006, "Galactic Gradients, Postbiological Evolution and the Apparent Failure of SETI," *New Ast.* **11**, 628-639.
[10.] Kardashev, N. S. 1964, "Transmission of information by extraterrestrial civilizations," *Sov. Astron.* **8**, 217-220.
[11.] Stapledon, O. 1930, *Last and First Men* (Millenium, London).
[12.] Bostrom, N. 2005, "A History of Transhumanist Thought," *Journal of Evolution and Technology*, **14**, No. 1 (http://www.nickbostrom.com/papers/history.pdf).
[13.] Bostrom, N. 2008, "The Future of Humanity," in *New Waves in Philosophy of Technology*, eds. J. B. Olsen, E. Selinger & S. Riis (Palgrave, McMillan), in press.
[14.] Kurzweil, R. 1999, *The Age of Spiritual Machines: When computers exceed human intelligence* (Viking, New York).
[15.] Kurzweil, R. 2005, *The Singularity is Near* (Duckworth, London).
[16.] Ćirković, M. M. 2003, "On the Importance of SETI for Transhumanism," *Journal of Evolution and Technology*, vol. **13** (http://www.jetpress.org/volume13/cirkovic.html).
[17.] Hanson, R. 1999, "Great Filter" (preprint at http://hanson.berkeley.edu/greatfilter.html).
[18.] Bostrom, N. 2002, "Existential Risks," *Journal of Evolution and Technology* **9** (http://www.jetpress.org/volume9/risks.html).
[19.] Dyson, F. J. 1979, "Time without end: Physics and biology in an open universe," *Reviews of Modern Physics* **51**, 447-460.
[20.] Adams, F. C. and Laughlin, G. 1997, "A dying universe: the long-term fate and evolution of astrophysical objects," *Rev. Mod. Phys.* **69**, 337-372.

[21.] Ćirković, M. M. 2003, "Resource Letter PEs-1: Physical eschatology," *Am. J. Phys.* **71**, 122-133.
[22.] Bostrom, N. 2001, "The Doomsday Argument, Adam & Eve, UN$^{++}$ and Quantum Joe," *Synthese* **127**, 359.
[23.] Sandberg, A. 1999, "The physics of information processing superobjects: daily life among the Jupiter brains," *J. Evol. Tech.* **5** (http://www.jetpress.org/volume5/Brains2.pdf).
[24.] Tinbergen, N. 1968, "On war and peace in animals and man," *Science* **160**, 1411-1418.
[25.] Tiger, L. and Fox, R. 1971, *The Imperial Animal* (Holt, Rinehart & Winston, New York).
[26.] Wilson, E. O. 1978, *On Human Nature* (Harvard University Press, Cambridge).
[27.] Trivers, R. L. 1985, *The Social Evolution* (Benjamin Cummings, Menlo Park).
[28.] Bostrom, N. 2003, "Astronomical Waste: The Opportunity Cost of Delayed Technological Development," *Utilitas* 15, 308-314.
[29.] Parkinson, B. 2005, "The carbon or silicon colonization of the universe?" *J. Brit. Interplan. Soc.* **58**, 111-116.
[30.] Rose, C. and Wright, G. 2004, "Inscribed matter as an energy-efficient means of communication with an extraterrestrial civilization," *Nature* **431**, 47-49.
[31.] Richmond, J. A. 1998, "Spies in Ancient Greece," *Greece & Rome* **45**, 1-18.
[32.] Dick, S. J. 2003, "Cultural Evolution, the Postbiological Universe and SETI," *Int. J. Astrobiology* **2**, 65-74.
[33.] Moravec, H. 1989, *Mind Children* (Harvard University Press, Cambridge).
[34.] Moravec, H. 1999, *Robot: Mere Machine to Transcendent Mind*, (Oxford University Press, New York).
[35.] Egan, G. 1997, *Diaspora* (Orion/Millennium, London).
[36.] Hammond, M. 1948, "Ancient Imperialism: Contemporary Justifications," *Harvard Studies in Classical Philology* **58**, 105-161.
[37.] Zuckerman, B. 1985, "Stellar Evolution: Motivation for Mass Interstellar Migrations," *Q. Jl. R. astr. Soc.* **26**, 56-59.
[38.] Criswell, D. R. 1985, "Solar System Industrialization: Implications for Interstellar Migrations," in R. Finney and E.M. Jones (Eds), *Interstellar Migration and the Human Experience* (University of California Press, Berkeley), 50-87.
[39.] Beech, M. 2008, *Rejuvenating the Sun and Avoiding Other Global Catastrophes* (Springer, New York).
[40.] Beech, M. 1990, "Blue stragglers as indicators of extraterrestrial civilisations?" *Earth, Moon, and Planets* **49**, 177-186.
[41.] Clarence-Smith, W. G. 1979, "The Myth of Uneconomic Imperialism: The Portuguese in Angola 1836-1926," *Journal of Southern African Studies* **5**, 165-180.
[42.] Kennedy, P. 1987, *The Rise and Fall of the Great Powers* (Random House, London).
[43.] Kennedy, P. 1989, "The Costs and Benefits of British Imperialism 1846-1914," *Past and Present* **125**, 186-192.
[44.] Hanson, R. 1998, "Burning the Cosmic Commons: Evolutionary Strategies for Interstellar Colonization" (preprint at http://hanson.gmu.edu/filluniv.pdf).
[45.] Ward, P. D. and Brownlee, D. 2000, *Rare Earth: Why Complex Life Is Uncommon in the Universe* (Springer, New York).

[46.] Rummel, J. D. 2001, "Planetary exploration in the time of astrobiology: Protecting against biological contamination," *Publications of the National Academy of Science* **98**, 2128-2131.

[47.] Grinspoon, D. 2003, *Lonely Planets: The Natural Philosophy of Alien Life* (HarperCollins, New York).

[48.] Cornford, F. 1965, *Thucydides Mythistoricus* (Greenwood Press Publishers, New York).

[49.] Kagan, D. 1969, *The Outbreak of the Peloponnesian War* (Cornell University Press, Ithaca).

[50.] Rees, M. J. 2003, *Our Final Hour: A Scientist's Warning: How Terror, Error, and Environmental Disaster Threaten Humankind's Future In This Century—On Earth and Beyond* (Basic Books, New York).

[51.] Caplan, B. 2008, "The Totalitarian Threat," in *Global Catastrophic Risks* by N. Bostrom and M. M. Ćirković (eds.) (Oxford University Press, Oxford, 2008), in press.

[52.] Hobson, J. A. 1902, *Imperialism: A Study* (Cosimo, London).

[53.] Wright, J. C. 2002, *The Golden Age* (Tor, New York).

[54.] Wright, J. C. 2003, *The Phoenix Exultant* (Tor, New York).

[55.] Wright, J. C. 2003, *The Golden Transcendence* (Tor, New York).

[56.] Brin, G. D. 1983, "The great silence – the controversy concerning extraterrestrial intelligent life," *Q. Jl. R. astr. Soc.* **24**, 283-309.

[57.] Webb, S. 2002, *Where is Everybody? Fifty Solutions to the Fermi's Paradox* (Copernicus, New York).

[58.] Lineweaver, C. H., Fenner, Y., and Gibson, B. K. 2004, "The Galactic Habitable Zone and the Age Distribution of Complex Life in the Milky Way," *Science* **303**, 59-62.

[59.] Olum, K. 2004, "Conflict between anthropic reasoning and observation," *Analysis* **64**, 1-8.

[60.] Lineweaver, C. H. 2001, "An Estimate of the Age Distribution of Terrestrial Planets in the Universe: Quantifying Metallicity as a Selection Effect," *Icarus* **151**, 307-313.

[61.] Ball, J. A. 1973, "The Zoo Hypothesis," *Icarus*, **19**, 347-349.

[62.] Fogg, M. J. 1987, "Temporal Aspects of the Interaction among the First Galactic Civilizations: The 'Interdict Hypothesis'," *Icarus* **69**, 370-384.

[63.] Baxter, S. 2000, "The planetarium hypothesis: a resolution of the Fermi paradox," *J. Br. Interplan. Soc.* **54**, 210-216.

[64.] Bostrom, N. 2003. "Are You Living In A Computer Simulation?" *Phil. Quart.* **53**, pp. 243-255.

[65.] Darling, D. 2001, *Life Everywhere* (Basic Books, New York).

[66.] Carter, B. 1983, "The anthropic principle and its implications for biological evolution," *Philos. Trans. R. Soc. London A* **310**, 347-363.

[67.] Ćirković, M. M. 2004, "Earths: Rare in Time, Not Space?" *J. Br. Interplan. Soc.* **57**, 53-59.

[68.] Gros, C. 2005, "Expanding Advanced Civilizations in the Universe," *J. Br. Interplan. Soc.* **58**, 108-111.

[69.] Lem, S. 1977, *Summa Technologiae* (Nolit, Belgrade).

[70.] Wesson, P. S. 1990, "Cosmology, Extraterrestrial Intelligence, and a Resolution of the Fermi-Hart Paradox," *Q. Jl. R. astr. Soc.* **31**, 161-170.

[71.] Annis, J. 1999, "Placing a limit on star-fed Kardashev type III civilizations," *J. Brit. Interplan. Soc.* **52**, 33-36.

[72.] Ćirković, M. M. 2006, "Too Early? On the Apparent Conflict of Astrobiology and Cosmology," *Biology and Philosophy* **21**, 369-379.
[73.] Toynbee, A. J. 1974, *The Study of History – Abridgement of Volumes I-VI by D. C. Somervell* (Oxford University Press, Oxford).
[74.] Hume, D. 1742, *Of the Rise and Progress of the Arts and Sciences* (available at http://infomotions.com/etexts/philosophy/1700-1799/hume-of-737.htm).
[75.] Levinson P. 2003, *REAL SPACE: The fate of physical presence in the digital age, on and off planet* (Routledge, London).
[76.] Krauss, L. M. and Starkman, G. D. 2000, "Life, The Universe, and Nothing: Life and Death in an Ever-Expanding Universe," *Astrophysical Journal* **531**, 22-30.
[77.] Parkinson, B. 2004, "Thoughts on the Evolution of Consciousness," *J. Br. Interplan. Soc.* **57**, 60-66.